\documentclass[10pt,twocolumn,journal]{IEEEtran}
\usepackage{cite}
\usepackage{graphicx}
\usepackage{subfigure}
\usepackage{algorithm,algorithmic}
\usepackage{threeparttable}
\usepackage{array}
\usepackage{amsmath}
\usepackage{amsfonts,amssymb}
\usepackage{stfloats}
\usepackage{bm}

\renewcommand{\baselinestretch}{.8}

\begin{document}
\title{Gibbs Distribution Based Antenna Splitting and User Scheduling in Full Duplex Massive MIMO Systems}

\author{\IEEEauthorblockN{Mangqing Guo and M. Cenk Gursoy} \\
\thanks{Copyright (c) 2015 IEEE. Personal use of this material is permitted. However, permission to use this material for any other purposes must be obtained from the IEEE by sending a request to pubs-permissions@ieee.org.
	
The authors are with the Department of Electrical Engineering and Computer Science, Syracuse University, Syracuse, NY 13244.
(Email: mguo06@syr.edu, mcgursoy@syr.edu)}
}
\maketitle

\begin{abstract}
A Gibbs distribution based combinatorial optimization algorithm for joint antenna splitting and user scheduling problem in full duplex massive multiple-input multiple-output (MIMO) system is proposed in this paper. The optimal solution of this problem can be determined by exhaustive search. However, the complexity of this approach becomes prohibitive in practice when the sample space is large, which is usually the case in massive MIMO systems. Our algorithm overcomes this drawback by converting the original problem into a Kullback-Leibler (KL) divergence minimization problem, and solving it through a related dynamical system via a stochastic gradient descent method. Using this approach, we improve the spectral efficiency (SE) of the system by performing joint antenna splitting and user scheduling. Additionally, numerical results show that the SE curves obtained with our proposed algorithm overlap with the curves achieved by exhaustive search for user scheduling.
\end{abstract}

\section{Introduction}
Supporting the fast increasing demand for data traffic is one of the core tasks in 5G or beyond 5G wireless communication systems, and several advanced technologies, including massive multiple-input multiple-output (MIMO), millimeter wave transmissions, full-duplex operation, and device-to-device (D2D) communication, have been proposed and analyzed intensively in recent years. Above all, compared to traditional half duplex systems, full duplex communication has the potential of doubling the spectral efficiency (SE) of the system by supporting simultaneous uplink and downlink transmissions on the same frequency band. However, which antenna at the base station (BS) should work in uplink mode, and which should work in downlink mode? Can we improve the SE of the system by suitable antenna splitting at the BS? There are generally large number of users in massive MIMO systems. Some users require transmission to the BS, while other users need to receive messages from the BS. Since the capacity of the system is limited, only a subset of these users could be served simultaneously. Thus, which user should be served at a given time instant? Can we further improve the SE of the system with an appropriate user scheduling scheme?

Answering the above questions is critical and indeed we can improve the SE of the system with suitable antenna splitting and user scheduling schemes. Specifically, our work in this paper addresses how to approach the optimal joint antenna splitting and user scheduling scheme that maximizes the SE of full duplex single-cell massive MIMO systems. We note that there already are several existing studies considering either antenna splitting or user scheduling in full duplex single-cell massive MIMO communication systems. For instance, an iterative convex optimization and binary relaxation based method for uplink and downlink antenna splitting has been proposed in \cite{Silva2018}. Antenna selection in full duplex cooperative non-orthogonal multiple access (NOMA) systems is considered in \cite{YuChenLiEtAl2018}. The authors in \cite{Ahn2016} propose a successive user selection algorithm to maximize the SE of the system. The user selection algorithm proposed in \cite{YuKohKang2017} considers the tradeoff between increasing the SE and decreasing the user-to-user interference. However, to the best of our knowledge, there is no work in the literature that focuses on joint antenna splitting and user scheduling in full duplex massive MIMO communication systems, which can further improve the SE of the system.

The Gibbs distribution based statistical combinatorial optimization algorithm proposed in this paper provides an efficient way to approach the optimal solutions of the joint antenna splitting and user scheduling problem in full duplex single-cell massive MIMO communication systems. This algorithm converts the original combinatorial optimization problem into Kullback-Leibler (KL) divergence minimization, and then solves it within a stochastic dynamical system.

The remainder of the paper is organized as follows. In Section \ref{sec:model}, we present the system model. In Section \ref{sec:learning-based}, we describe the proposed Gibbs distribution based combinatorial optimization algorithm for joint antenna splitting and user scheduling. Numerical results are provided in Section \ref{sec:numerical} and conclusions are drawn in Section \ref{sec:conclusion}.

\section{System Model} \label{sec:model}
\begin{figure}[htbp]
  \centering
  \includegraphics[width=2.5in]{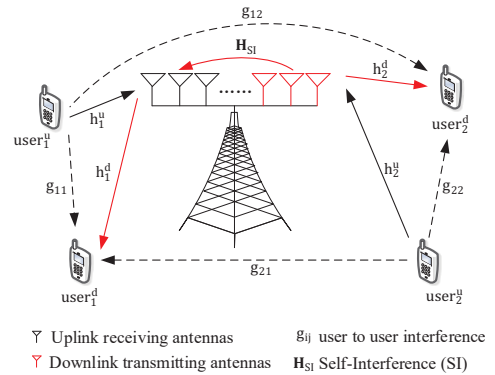}\\
  \caption{Full duplex single-cell Massive MIMO system.}\label{fig1}
\end{figure}

Consider a single-cell massive MIMO system equipped with one BS, a set of antennas denoted by ${\cal M}$ and a set of single-antenna users, denoted by ${\cal N}$. The number of antennas and users are $\left| {\cal M} \right| = M$, $\left| {\cal N} \right| = N$, respectively, where $\left|  \cdot  \right|$ denotes the cardinality of a given set. The BS operates in full duplex mode. Each antenna at the BS is connected to two analog circuits, and can be configured to work in uplink or downlink mode. A subset of users requires transmission to the BS, while the others need to receive data from the BS. The requirements of the users vary dynamically. As shown in Fig. \ref{fig1}, the transmitted signals from the uplink users will cause interference to downlink users, and the transmission from the downlink transmit antennas will introduce self-interference (SI) to uplink receive antennas.

At a given time instant, we denote the set of users which need to transmit messages to the BS as ${{\cal K}_u}$, while the set of users which require to receive data from the BS as ${{\cal K}_d}$. We consider flat fading channel and assume transmission over only a single frequency band in this paper. We assume that the activated sets of uplink users, downlink users, transmitting antennas, and receiving antennas are denoted by ${{\cal N}_{\text{u}}}$, ${{\cal N}_{\text{d}}}$, ${{\cal M}_{\text{t}}}$ and ${{\cal M}_{\text{r}}}$, respectively. We have the following relationships among these sets:
\begin{IEEEeqnarray}{rCl}\label{equadd0}
&& {{\cal M}_{\text{t}}} \cup {{\cal M}_{\text{r}}} = {\cal M},{\kern 5pt}{{\cal M}_{\text{t}}} \cap {{\cal M}_{\text{r}}} = \emptyset ,{\kern 5pt}{{\cal N}_{\text{u}}} \subset {{\cal K}_{\text{u}}} \subset {\cal N}, \nonumber\\
&& {{\cal N}_{\text{d}}} \subset {{\cal K}_{\text{d}}} \subset {\cal N},{\kern 5pt}{{\cal N}_{\text{u}}} \cap {{\cal N}_{\text{d}}} = \emptyset.
\end{IEEEeqnarray}
Moreover, we denote the cardinalities as $\left| {{{\cal N}_{\text{u}}}} \right| = {N_{\text{u}}}$, $\left| {{{\cal N}_{\text{d}}}} \right| = {N_{\text{d}}}$, $\left| {{{\cal M}_{\text{t}}}} \right| = {M_{\text{t}}}$ and $\left| {{{\cal M}_{\text{r}}}} \right| = {M_{\text{r}}}$.

As also depicted in Fig. \ref{fig1}, ${\mathbf{h}}_i^u \in {\mathcal{C}^{{M_r} \times 1}}$ and ${\mathbf{h}}_j^d \in {\mathcal{C}^{1 \times {M_t}}}$ are the channel vectors for uplink and downlink transmissions, respectively. $g_{ij}$ is the interference channel coefficient from $i$th uplink user to $j$th downlink user. Flat fading is assumed in uplink, downlink and user-to-user interference channels, and pathloss, shadowing and small scale fading are taken into account. The pathloss model in 3GPP TR 36.828 \cite{3GPP2012} is employed in this paper.
The shadow fading is assumed to have a log-normal distribution with standard deviation ${\sigma _{{\rm{shadow}}}}$. The small-scale fading components are modeled as independent identically distributed (i.i.d.) complex Gaussian random variables with zero mean and unit variance. ${{\bf{H}}_{{\text{SI}}}}\sim {{\mathcal C}^{{M_r} \times {M_t}}}$ is the residual SI matrix from the transmit antennas to receive antennas at the BS with entries that are determined by the SI cancellation techniques. In particular, ${{\bf{H}}_{{\text{SI}}}}$ can be characterized to have non-zero mean complex Gaussian components, i.e., ${{\bf{H}}_{{\text{SI}}}} \sim {\cal C}{\cal N}(\sqrt {\sigma _{{\text{SI}}}^2\kappa /(1 + \kappa )} {\overline {\bf{H}} _{{\text{SI}}}},(\sigma _{{\text{SI}}}^2/(1 + \kappa )){{\bf{I}}_{{N_r}}} \otimes {{\bf{I}}_{{N_t}}})$ where ${\overline {\bf{H}} _{{\text{SI}}}}$ is a constant matrix, $\kappa $ denotes the Rician factor, $\sigma _{{\text{SI}}}^2$ represents the SI power, and $ \otimes $ stands for the Kronecker product of two matrices \cite{Ahn2016}\cite{Duarte2012}. We assume that the channel state information of all links is known at the BS.

The received signal at the $k$th downlink user is
\begin{equation}\label{equ1}
y_k^{\rm{d}} = \sqrt {{p_{\rm{d}}}} {\bf{h}}_k^{\rm{d}}{\bf{Wz}} + \sum\nolimits_{j = 1}^{{N_{\rm{u}}}} {\sqrt {{p_{\rm{u}}}} {g_{kj}}{s_j}}  + n_k^{\rm{d}}
\end{equation}
where ${{p_{\rm{u}}}}$ and ${{p_{\rm{d}}}}$ are the uplink and downlink transmit power levels, respectively. ${\bf{W}}$ is the precoding matrix for the downlink transmit signals at the BS. ${\bf{z}}$ is an ${N_{\rm{d}}} \times 1$-dimensional vector which comprises the message transmitted from the BS to all the downlink users, satisfying $\mathbb{E}\{ {\left| {{z_i}} \right|^2}\}  = 1$, where ${{z_i}}$ is the $i$th element of ${\bf{z}}$. ${{s_j}}$ is the data signal transmitted from the $j$th user to the BS, and this signal also satisfies $\mathbb{E}\{ {\left| {{s_j}} \right|^2}\}  = 1$. $n_k^{\text{d}} \sim \mathcal{C}\mathcal{N}(0,{(\sigma _k^{\text{d}})^2})$ is additive white Gaussian noise (AWGN) at the $k$th downlink user.

Denote by ${{\bf{H}}_{\text{d}}} = {[{({\bf{h}}_1^{\text{d}})^{\text{H}}},{({\bf{h}}_2^{\text{d}})^{\text{H}}},...,{({\bf{h}}_{{N_{\text{d}}}}^{\text{d}})^{\text{H}}}]^{\text{H}}}$ the channel matrix from all the downlink transmit antennas to the downlink users. We use zero-forcing (ZF) precoding at the BS for downlink data transmission to eliminate multiuser interference. Let us define ${{\bf{F}}_{\text{d}}} = {\bf{H}}_{\text{d}}^{\text{H}}{({{\bf{H}}_{\text{d}}}{\bf{H}}_{\text{d}}^{\text{H}})^{ - 1}}$. Then, the normalized ZF precoding matrix for the downlink transmitted signals at the BS is
\begin{equation}\label{equ3}
  {\bf{W}} = \frac{{{{\bf{F}}_{\text{d}}}}}{{{{\left\| {{{\bf{F}}_{\text{d}}}} \right\|}_{\text{F}}}}}
\end{equation}
where ${\left\| {{{\bf{F}}_{\text{d}}}} \right\|_{\text{F}}}$ is the Frobenius norm of matrix ${{{\bf{F}}_{\text{d}}}}$. The normalization of the ZF precoding matrix is used to control the transmission power of the signals from the BS to all the downlink users, to make the total downlink transmission power equal to $p_d$.

Now, the received signal-to-interference-plus-noise ratio (SINR) of the $k$th downlink user is
\begin{equation}\label{equ4}
\gamma _k^{\text{d}} = \frac{{{p_{\text{d}}}{{\left| {{\bf{h}}_k^{\text{d}}{{\bf{w}}_k}} \right|}^2}}}{{\sum\nolimits_{j = 1}^{{N_u}} {{p_{\text{u}}}{{\left| {{g_{kj}}} \right|}^2}}  + {{(\sigma _k^{\text{d}})}^2}}}
\end{equation}
where ${{\bf{w}}_i}$ is the $i$th column of ${\bf{W}}$.

The received uplink signal at the BS is
\begin{IEEEeqnarray}{rCl}\label{equ5}
  {{\bf{y}}^{\text{u}}} = \sqrt {{p_{\text{u}}}} {{\bf{H}}_{\text{u}}}{\bf{s}} + \sqrt {p_{\text{d}}} {{\bf{H}}_{{\text{SI}}}}{\bf{Wz}} + {{\bf{n}}_{\text{u}}}
\end{IEEEeqnarray}
where ${{\bf{H}}_{\bf{u}}} = [{\bf{h}}_1^{\text{u}},{\bf{h}}_2^{\text{u}},...,{\bf{h}}_{{N_{\text{u}}}}^{\text{u}}]$ is the channel matrix from all uplink users to all receiving antennas at the BS. ZF reception is used at the BS, and the received noisy signal corresponding to the signal transmitted from the $k$th uplink user is
\begin{equation}\label{equ6}
  \widehat y_k^{\text{u}} = \sqrt {{p_{\text{u}}}} {{\bf{p}}_k}{{\bf{H}}_{\text{u}}}{\bf{s}} + \sqrt {{p_{\text{d}}}} {{\bf{p}}_k}{{\bf{H}}_{{\text{SI}}}}{\bf{Wz}} + {{\bf{p}}_k}{{\bf{n}}_{\text{u}}}
\end{equation}
where ${\bf{P}} = {({\bf{H}}_{\text{u}}^{\text{H}}{{\bf{H}}_{\text{u}}})^{ - 1}}{\bf{H}}_{\text{u}}^{\text{H}}$ is the ZF detection matrix for the uplink signals, and ${{{\bf{p}}_k}}$ is the $k$th row of ${\bf{P}}$. Then, the SINR of the received signal transmitted from the $k$th uplink user is
\begin{equation}\label{equ7}
\gamma _k^{\text{u}} = \frac{{{p_{\text{u}}}{\text{ }}}}{{{p_{\text{d}}}\sum\limits_{i = 1}^{{N_{\text{d}}}} {{{\left| {{{\bf{p}}_k}{{\bf{H}}_{{\text{SI}}}}{{\bf{w}}_i}} \right|}^2}}  + {\sigma ^2}\left\| {{{\bf{p}}_k}} \right\|_2^2}}
\end{equation}
in which ${{\sigma ^2}}$ is the AWGN noise variance at the BS.

Given these assumptions and characterizations, the achievable SE of the system in bps/Hz is
\begin{equation}\label{equ8}
R = \sum\limits_{i = 1}^{{N_{\text{u}}}} {{{\log }_2}(1 + \gamma _i^{\text{u}})}  + \sum\limits_{j = 1}^{{N_{\text{d}}}} {{{\log }_2}(1 + \gamma _j^{\text{d}})}.
\end{equation}

Our primary goal is to find a joint antenna splitting and user scheduling scheme that maximizes the achievable $R$ in (\ref{equ8}), i.e., solves the following optimization problem:
\begin{IEEEeqnarray}{rCl}\label{equ9}
   \mathop {\max }\limits_{{\cal M}_{\text{r}}^ 1 ,{\cal M}_{\text{t}}^ 1 ,{\cal K}_{\text{u}}^1 ,{\cal K}_{\text{d}}^ 1 }{\kern 10pt} && R  \IEEEyesnumber\IEEEyessubnumber*\\
   {\text{subject}}{\kern 5pt}{\text{to}}{\kern 10pt}&& K_{\min }^{\text{u}} \leqslant \left| {{\cal K}_{\text{u}}^1 } \right| \leqslant \left| {{\cal M}_{\text{r}}^ 1 } \right|  \IEEEyessubnumber\\
  && K_{\min }^{\text{d}} \leqslant\left| {{\cal K}_{\text{d}}^ 1 } \right| \leqslant \left| {{\cal M}_{\text{t}}^1 } \right|  \IEEEyessubnumber
\end{IEEEeqnarray}
where ${\cal M}_{\text{r}}^ 1$, ${\cal M}_{\text{t}}^ 1$, ${\cal K}_{\text{u}}^ 1$, ${\cal K}_{\text{d}}^ 1$ are the subsets of uplink and downlink BS antennas and users, respectively, and they should satisfy the constraints in (\ref{equadd0}). In order to achieve the diversity gain of a MIMO system, the upper bound constraints on $\left| {{\cal K}_{\text{u}}^1 } \right|$ and $\left| {{\cal K}_{\text{d}}^ 1 } \right|$ in (9b) and (9c), respectively, should be satisfied. The lower bound constraints in (9b) and (9c) ensure the fairness in uplink and downlink transmissions in full duplex communication systems.

Problem (\ref{equ9}) is a combinatorial optimization problem, and it is rather difficult to get an analytical solution. In general, we can obtain the optimal solution of this combinatorial optimization problem via exhaustive search. However, when the sample space is large (as is often the case in massive MIMO systems), it becomes prohibitive in complexity to use exhaustive search in practice. On the other hand, the Gibbs distribution based statistical combinatorial optimization algorithm introduced in the following section provides an effective method to address this problem with lower complexity.

\section{Gibbs Distribution Based Statistical Combinatorial Optimization Algorithm} \label{sec:learning-based}
In this section, we analyze how to solve combinatorial optimization problems using the statistical framework. For completeness in the paper, we initially review the original statistical framework proposed in \cite{Berny2000}, and then address its advantages and disadvantages in solving the problem in (\ref{equ9}). Following this, we provide several extensions to the original statistical framework to make it an efficient algorithm for solving the problem in (\ref{equ9}).
\subsection{Statistical Framework}
Suppose we have a combinatorial optimization problem with $n$ features ${\bf{x}} = {[{x_1},{x_2},...,{x_n}]^T}$, ${x_i} \in \{ 0,1\} $, and we want to minimize the objective function $f(\bf{x})$. In \cite{Berny2000}, a statistical framework is proposed for solving this problem based on the characteristics of Gibbs distribution and dynamical systems.

The Gibbs distribution maps each value of the objective function onto a probability defined by
\begin{equation}\label{equg1}
 p_T^ * ({\bf{x}}) = {{\exp ( - f({\bf{x}})/T)} \over {\sum\nolimits_{{\bf{y}} \in {\cal S}} {\exp ( - f({\bf{y}})/T)} }}
\end{equation}
where $T>0$ is the analogue of a temperature\footnote{In statistical mechanics, a Gibbs distribution is a probability distribution that gives the probability that a system will be in a certain state as a function of that state's energy and the temperature of the system. In the combinatorial optimization problem, we can regard different inputs as different states of a system, and the objective function value as the corresponding state's energy. Then, a set of Gibbs distributions with different temperatures, $T$, corresponds to a set of different distributions of the states of the system. The convergence property of the Gibbs distribution which will be used later to solve the combinatorial optimization problem depends on the choice of the temperature $T$.} and ${\cal S} = {\{ 0,1\} ^n}$ is the set of all possible values of ${\bf{x}}$. We denote ${\cal U} = \{ {\bf{x}} \in {\cal S}|\forall {\bf{y}} \in {\cal S},f({\bf{x}}) \le f({\bf{y}})\} $. Then, the Gibbs distribution $p_T^ * $ converges to a uniform distribution on ${\cal U}$ when $T$ tends to zero. In other words, we can obtain the optimal solutions by finding the limit distribution of $p_T^ * $ as $T \to 0$. However, this is difficult in practice. So instead of finding the limit distribution directly, we search for a distribution which has the smallest KL divergence to an implicit Gibbs distribution. The KL divergence between $p$ and $p_T^ * $ is
\begin{equation}\label{equg2}
  D(p,p_T^ * ) =  - \sum\limits_{{\bf{x}} \in {\cal S}} {p({\bf{x}}) \ln{{p_T^ * ({\bf{x}})} \over {p({\bf{x}})}}}
\end{equation}
where $\ln( \cdot )$ is the natural logarithm function. Then, the problem is converted into the minimization of the free energy of the system:
\begin{equation}\label{equg3}
  F = \sum\nolimits_{{\bf{x}} \in {\cal S}} {p({\bf{x}})f({\bf{x}}) + T\sum\nolimits_{{\bf{x}} \in {\cal S}} {p({\bf{x}}){\ln}(p({\bf{x}}))} } .
\end{equation}

This is still a discrete problem, and it is not easy to solve in practice. By introducing ${\bm{\theta }} = {[{\theta _1},{\theta _2},...,{\theta _n}]^T}$ as the probability distribution parameter for the $n$-dimensional random vector ${\bf{x}}={[{x _1},{x _2},...,{x _n}]^T}$, we can convert
the discrete optimization into a continuous optimization problem. Then, we introduce the following dynamical system
\begin{equation}\label{equg4}
  \frac{{\partial {\bm{\theta }}}}{{\partial t}} + \frac{{\partial F}}{{\partial {\bm{\theta }}}} = 0.
\end{equation}
which is used to approach the minimum value of the system's free energy $F$. Then, with the stochastic approximation technique, we obtain the following statistical update rule:
\begin{equation}\label{equ45}
{\theta _i}(t + 1) = {\theta _i}(t) - \alpha \left(f({\bf{x}}) + T(1 + \ln(p({\bf{x}})))\right){{\partial \ln(p({\bf{x}}))} \over {\partial {\theta _i}}}
\end{equation}
where $\alpha$ is a preassigned constant, and ${{\partial \ln(p({\bf{x}}))} \over {\partial {\theta _i}}}$ is the gradient \cite{Berny2000}.

Suppose we choose the random vector ${\bf{x}}$ to be multivariate Bernoulli distributed, and all the elements are independent of each other. Then, the joint distribution of $\bf{x}$ is as follows:
\begin{equation}\label{equ46}
  p({\bf{x}}) = \prod\limits_{i = 1}^n {({x_i}{p _i} + (1 - {x_i})(1 - {p_i}))}
\end{equation}
where $p_i$ is the probability that $x_i$ equals to 1. The relationship between $p_i$ and $\theta_i$ is
\begin{equation}\label{equ47}
  {p_i} = \frac{1}{2}(1 + \tanh (\beta {\theta _i})),
\end{equation}
where $\beta$ is the sigmoid gain, which is used to control the gradient of the multivariate Bernoulli distribution. The gradient of $\ln(p({\bf{x}}))$ is
\begin{equation}\label{equ48}
  {{\partial \ln(p({\bf{x}}))} \over {\partial {\theta _i}}} = 2\beta ({x_i} - {p_i}).
\end{equation}
With this, the update rule in (\ref{equ45}) becomes
\begin{equation}\label{equup}
  {\theta _i}(t + 1) = {\theta _i}(t) - 2\alpha \beta \left( {f({\bf{x}}) + T(1 + \ln(p({\bf{x}})))} \right)({x_i} - {p_i}),
\end{equation}
where $\alpha$ and $\beta$ are preassigned constants of this statistical framework.

The parameters $\alpha$ and $\beta$ are selected empirically in this paper. Numerical simulations show that we do not need to change $\alpha$ and $\beta$ when the channel coefficients change. We set $\alpha=0.5$, $\beta=0.2$ (when $SNR\leqslant 10dB$) and $\beta=0.1$ (when $SNR>10dB$) during the numerical simulations in Section \uppercase\expandafter{\romannumeral4}.

As noted before, the Gibbs distribution converges to a uniform distribution which only has non-zero values on the optimal solutions. If we use the Metropolis algorithm to update the Gibbs distribution to get the optimal solutions, then we have the well-known simulated annealing algorithm. As the objective function decreases fastest in the gradient direction, this statistical optimization algorithm will arrive at the optimal solution quicker than simulated annealing.

If we are interested in finding the maximum value of an objective function, we just need to add a negative sign before the objective function, and then substitute it into our algorithm.

Below, we provide the Gibbs distribution based statistical combinatorial optimization algorithm using the gradient descent method.
\begin{algorithm}
 \caption{Gibbs distribution based statistical combinatorial optimization algorithm.}
 \label{alg0}
 \begin{algorithmic}[1]
  \STATE Initialize the multivariate Bernoulli distribution parameter $\bm{\theta}$, learning rate $\alpha$, and $\beta$;
  \STATE Generate an $n$-dimensional multivariate Bernoulli random vector $\textbf{x}$ which satisfies all the constraints in the combinatorial optimization problem, with parameter vector $\bm{\theta}$;
  \STATE Calculate the objective function value $f(\textbf{x})$;
  \STATE Update $\bm{\theta}$ with (\ref{equup});
  \STATE If the stop criteria is met, stop; otherwise, go to step 2.
 \end{algorithmic}
\end{algorithm}
\subsection{Drawbacks of Algorithm \ref{alg0} and the Corresponding Solutions}
Although Algorithm \ref{alg0} could arrive at the maximum value of the objective function, it has fluctuations after getting the optimum solution, which makes it difficult for us to recognize the optimal solution. In practice, we would like the algorithm to converge and stay at the optimum value.

Another drawback lies in step 2 of Algorithm \ref{alg0}, which should generate an $n$-dimensional multivariate Bernoulli random vector $\bf{x}$ which satisfies all the constraints of the combinatorial optimization problem with parameter vector $\bm{\theta}$. It may become difficult to produce this $n$-dimensional multivariate Bernoulli random vector directly as the constraints for the combinatorial optimization problem become strict, e.g., especially when the probability of the $n$-dimensional multivariate Bernoulli random vector, $p({\bf{x}})$, is very small, such as less than ${10^{ - 10}}$. Such low probabilities may be experienced in practice and such cases require a large number of samples to get a realization of the event, which is impractical in real full duplex massive MIMO systems.

These observations motivate us to introduce the following modifications to Algorithm 1 to overcome the drawbacks:
\subsubsection{Convergence}
 The fluctuation problem of Algorithm \ref{alg0} could be solved by introducing some revisions in steps 2 and 3. Instead of generating only one sample, we generate a population of $N$ individuals from the $n$-dimensional multivariate Bernoulli distribution, which satisfy the constraints with parameter vector $\bm{\theta}$ in step 2. Then we determine the objective function value for each individual and select the one which leads to the smallest value in step 3. Numerical results demonstrate that Algorithm 1 converges to the optimal solutions quickly with these changes, and keeps constant after arriving at the optimum value.
\subsubsection{Rare event simulation}
The second drawback of Algorithm \ref{alg0} can be overcome by employing the efficient subset simulation method for rare event estimation, proposed by Au and Beck in \cite{Au2001}. The basic idea of subset simulation method is to decompose the rare event $F$ into a sequence of progressively ``less-rare" nested events $F = {F_L} \subset {F_{L - 1}} \subset  \cdots  \subset {F_1}$, where ${F_1}$ is a relatively frequent event \cite{Zuev2014}. Then the small probability $P(F)$ of the rare event $F$ can be represented as
\begin{equation}\label{equ50}
  P(F)=P(F_1)\cdot P(F_2|F_1)\cdot ... \cdot P(F_L|F_{L-1})
\end{equation}
where $P(F_k|F_{k-1})=P(F_k)/P(F_{k-1})$ is the conditional probability of $F_k$ given the occurrence of $F_{k-1}$, for $k=2,...,L$. With this, the estimation of rare event problem is transferred to the product of the probabilities of relatively more frequent events.

Rather than estimating the probability of a rare event, we could calculate the probability for each element of the $n$-dimensional multivariate Bernoulli random vector ${\bf{x}}$ which need to be generated. Our problem is to generate ${\bf{x}}$ with small probabilities without generating large amount of samples. However, during the process of estimating the probability of a rare event, random samples of this rare event are generated simultaneously. Thus, we can use the same process of subset simulation for rare event estimation to generate random samples of this rare event. The core problem of subset simulation is how to decompose the rare event into a sequence of relatively more frequent events. In practice, this is achieved by decomposing the rare event into a sequence of nested intermediate subsets ${F_k},k = 1,2, \cdots ,L$ with Markov Chain Monte Carlo (MCMC) technique, and the conditional probability satisfies $P(F_k|F_{k-1})=p_0$ (where $p_0=0.1$ is often used in the literature) \cite{Zuev2014}.

Suppose $F$ is a set of $M$-dimensional features $\textbf{z}$ which satisfy $h({\bf{z}}) \geqslant {\gamma ^ * }$, where $h( \cdot )$ is the objective function, and ${\gamma ^ * }$ is a given threshold. The procedure of subset simulation is a random sample generating process. At the beginning, $N$ samples $\{ {{\bf{z}}_n}\} $ $(n = 1,2, \cdots,N)$ are generated and arranged in decreasing order according to $h({{\bf{z}}_n})$, i.e., $h({{\bf{z}}_1}) \geqslant h({{\bf{z}}_2}) \geqslant  \cdots  \geqslant h({{\bf{z}}_N})$. Denote ${n_0} = \left\lceil {{p_0}N} \right\rceil $ as the smallest integer which is greater than or equal to ${p_0}N$. Since $F$ is a rare event, it is much more likely that $h({{\bf{z}}_1}) < {\gamma ^ * }$. However, if we arrange the entire feature set in decreasing order according to $h({\bf{z}})$, the samples with larger objective function values are closer to $F$ than the samples with smaller values, i.e., ${{\bf{z}}_1}$ is closet to $F$ while ${{\bf{z}}_N}$ is farthest to $F$. Therefore, we set ${\gamma _1} = h({{\bf{z}}_{{n_0}}})$, and denote the set of features which satisfy $h({\bf{z}}) \geqslant \gamma _1 $ as an intermediate subset $F_1$. It is obvious that $F \subset {F_1}$, and $F_1$ can be regarded as a conservative approximation to $F$. Then, based on $\{ {{\bf{z}}_n}\} $ $(n = 1,2, \cdots,{n_0})$, we generate another $N$ samples which belong to $F_1$ with the Modified Metropolis algorithm (MMA) \cite{Zuev2014} (the process of which is described later), and rearrange them in decreasing order according to $h({{\bf{z}}_n})$. We denote the number of samples which satisfy $h({{\bf{z}}_n}) \geqslant {\gamma ^ * }$ as $n_1$. If ${n_1} \geqslant {n_0}$, we stop the random sample generating process. Otherwise, we set another intermediate threshold $\gamma _2 $, determine the intermediate subset $F_2$ using the same method as when obtaining $\gamma _1 $ and $F_1$, and generate another $N$ samples based on the new set of samples $\{ {{\bf{z}}_n}\} $ $(n = 1,2, \cdots,{n_0})$ via the MMA algorithm and rearrange them, check the number of samples which belong to $F$, and make a decision according to whether ${n_2} \geqslant {n_0}$ or not. We repeat the process until the arrival to the point at which we have ${n_L} \geqslant {n_0}$ for the $L$th intermediate subset $F_L$. It is obvious that ${F_L} \subset {F_{L - 1}} \subset  \cdots  \subset {F_1}$, $P(F_2|F_1)=P(F_3|F_2) ... =P(F_L|F_{L-1})=\frac{n_o}{N}$, and $p(F) = \frac{{n_0^{L - 1}{n_L}}}{{{N^L}}}$.

The MMA algorithm is used to generate random samples which belong to an intermediate subset $F_k$. We consider the case in which each element of ${\bf{z}}$, e.g., ${z_m}$ $($for $m = 1,2, \cdots,M)$ is independent of other elements. In the procedure of subset simulation, for each intermediate subset $F_k$, we can always obtain a threshold ${\gamma _k}$ and subset of features $\{ {{\bf{z}}_n}\} $ $(n = 1,2, \cdots,{n_0})$. If ${n_k} < {n_0}$, another set of $N$ samples which belong to $F_k$ will be generated based on $\{ {{\bf{z}}_n}\} $ $(n = 1,2, \cdots,{n_0})$ with the MMA algorithm. $\{ {{\bf{z}}_n}\} $ $(n = 1,2, \cdots,{n_0})$ can be regarded as the seeds of the random samples to be generated. The process is as follows: (1) for each element of $\{ {{\bf{z}}_n}\} $ $(n = 1,2, \cdots,{n_0})$, ${z_{nm}}$ $(m = 1,2, \cdots,M)$, generate a corresponding uniform random variable ${\varsigma _{nm}}$ on $ [{z_{nm}}-2,{z_{nm}}+2] $; (2) calculate the acceptance ratio ${\omega _{nm}} = \frac{{{\varsigma _{nm}} - {z_{nm}} + 2}}{2}$; (3) accept ${{\varsigma _{nm}}}$ as the $m$th element of a candidate random sample with probability $\min \{ 1,{\omega _{nm}}\} $, and accept $ {z_{nm}} $ with probability $1-\min \{ 1,{\omega _{nm}}\} $; (4) accept the candidate random sample if it belongs to $F_k$; otherwise, reject this candidate random sample. The random samples generated with the MMA algorithm satisfy the conditional distribution of $F_{k+1}$ given $F_k$, and readers can refer to \cite{Zuev2014} for more details.

By now, random samples of the rare event, without any extra constraints, could be generated with the subset simulation method. However, the multivariate Bernoulli random vectors in this paper should satisfy the constraints in (\ref{equ9}), which can be achieved with the double-criterion ranking method introduced in\cite{Li2010}. For instance, if the constraints are $a \leqslant {\left\| {\bf{z}} \right\|_1} \leqslant b$, we define two constraint violation functions:
\begin{equation*}
{v_1}=
\begin{cases}
0,& if {\kern 5pt}{\left\| {\bf{z}} \right\|_1} \geqslant a \\
{\left\| {\bf{z}} \right\|_1}-a,& if {\kern 5pt}{\left\| {\bf{z}} \right\|_1} < a
\end{cases}
\end{equation*}
and
\begin{equation*}
{v_2}=
\begin{cases}
0,& if {\kern 5pt}{\left\| {\bf{z}} \right\|_1} \leqslant b \\
b-{\left\| {\bf{z}} \right\|_1},& if {\kern 5pt}{\left\| {\bf{z}} \right\|_1} > b.
\end{cases}
\end{equation*}
The overall constraint fitness function is defined as
\begin{equation*}
{v_{con}}({\bf{z}}) =  \min ({v_1},{v_2}).
\end{equation*}
In the subset simulation process, instead of rearranging $N$ random samples in decreasing order according to $h({{\bf{z}}_n})$, the double-criterion ranking method is employed with the following procedure. First, we rearrange the samples in decreasing order according to the values of ${v_{con}}({{\bf{z}}_n})$. Then, we reorder the samples which satisfy ${v_{con}}({{\bf{z}}_n})=0$ in decreasing order based on the values of $h({{\bf{z}}_n})$. The subset $F_k$ is defined as a set of samples which satisfy the following condition:
\begin{equation*}
{F_k}=
\begin{cases}
h({{\bf{z}}_n}) \geqslant h({{\bf{z}}_{{n_0}}}),& if {\kern 5pt} {v_{con}}({{\bf{z}}_n}) = 0 \\
{v_{con}}({{\bf{z}}_n}) \geqslant {v_{con}}({{\bf{z}}_{{n_0}}}),& if {\kern 5pt} {v_{con}}({{\bf{z}}_n}) < 0.
\end{cases}
\end{equation*}

We provide the process of generating multivariate Bernoulli random vectors which satisfy $a \leqslant {\left\| {\bf{z}} \right\|_1} \leqslant b$ with subset simulation method as Algorithm \ref{alg2} shown on the next page.
\begin{algorithm}
	\caption{$M$-dimensional Multivariate Bernoulli random vectors generating process under constraints $a \leqslant {\left\| {\bf{z}} \right\|_1} \leqslant b$ with subset simulation method.}
	\label{alg2}
	\begin{algorithmic}[1]
		\STATE Initialize the multivariate Bernoulli distribution parameter ${\bf{p^1 }}$ (the probability of ${z_m} = 1(m = 1,2, \cdots M)$ is $p_m^1$), the number of samples $N$, the iteration number $k=1$, the threshold ${\gamma ^ * }=0$ and the subset level probability $p_0$ for rare event simulation;
		\STATE Generate $N$ $M$-dimensional random vectors ${\bf{z}}_n^1(n = 1,2, \cdots N)$, where each element of ${\bf{z}}_n^1$ satisfies the uniform distribution on $[0,1]$.
		\STATE Denote the $m$th element of ${\bf{z}}_n^1$ and ${\bf{p^1}}$ as $z_{nm}^1$ and $p_m^1$, respectively. If $z_{nm}^1 \leqslant p_m^1$, ${z_{nm}}=1$; otherwise, ${z_{nm}}=0$.
		\STATE Count and denote the number of samples which satisfy ${v_{con}}({{\bf{z}}_n}) = 0$ as $n_1$.
		\WHILE{($n_k<n_0$)}		
		\STATE Find $h({{\bf{z}}_{{n_0}}})$ and ${v_{con}}({{\bf{z}}_{{n_0}}})$, and define the $k$th subset $F_k$.
		\STATE Generate another $N$ random samples which belong to $F_k$ with the MMA method, and convert them to multivariate Bernoulli vectors under the same way described in step 3.
		\STATE Count and denote the number of samples which satisfy ${v_{con}}({{\bf{z}}_n}) = 0$ as $n_k$.
		\IF{${n_k} \geqslant {n_0}$}
		\STATE break; \\
		\ELSE
		\STATE {k=k+1;}
	    \ENDIF
		\ENDWHILE		
	\end{algorithmic}
\end{algorithm}

As noted before, the subset simulation method will be employed in the second step of Algorithm \ref{alg0} to help generate $n$-dimensional multivariate Bernoulli random vectors with small probabilities, i.e., in the extended version of Algorithm \ref{alg0} (Algorithm \ref{alg1}) which will be described later, (in steps 2 and 3), we attempt to generate $N$ $n$-dimensional multivariate Bernoulli random vectors with parameter vector $\bm{\theta}$ first. Then, if none of the $N$ $n$-dimensional multivariate Bernoulli random vectors satisfies all the constraints in the combinatorial optimization problem, Algorithm \ref{alg2} is called in step 3 of Algorithm 3 to generate another $N$ multivariate Bernoulli random vectors with parameter vector $\bm{\theta}$ to replace the vectors generated with $\bm{\theta}$ directly in step 2 of Algorithm 3.

\subsection{Extended Gibbs Distribution Based Statistical Combinatorial Optimization Algorithm}
With the modifications introduced in the previous subsection, we provide the extended Gibbs distribution based statistical combinatorial optimization algorithm using the gradient descent method as Algorithm \ref{alg1} shown in next page.
\begin{algorithm}
 \caption{Extended Gibbs distribution based statistical combinatorial optimization algorithm.}
 \label{alg1}
 \begin{algorithmic}[1]
  \STATE Initialize the multivariate Bernoulli distribution parameter ${\bm{\theta }}$, preassigned parameters $\alpha$, $\beta$, and the number of samples at each iteration $N$;
  \STATE Generate $N$ $n$-dimensional multivariate Bernoulli random vectors $\textbf{x}$ with parameter vector $\bm{\theta}$;
  \STATE If none of the $N$ multivariate Bernoulli random vectors $\textbf{x}$ generated in step 2 satisfies all the constraints in the combinatorial optimization problem, Algorithm \ref{alg2} is used to produce another $N$ multivariate Bernoulli random vectors $\textbf{x}$, instead of the vectors produced in step 2.
  \STATE Calculate the objective function value for each of the multivariate Bernoulli random vectors $\textbf{x}$ generated in step 2 or 3, which satisfy all the constraints. Then, select the one that leads to the minimum objective function value;
  \STATE Update $\bm{\theta}$ with (\ref{equup}). As the goal is to maximize the SE in our problem, we substitute $f({\bf{x}}) =  - SE$ into (\ref{equup});
  \STATE If the stop criteria is met, stop; otherwise, go to step 2.
 \end{algorithmic}
\end{algorithm}

For solving the joint antenna splitting and user scheduling problem in (9) with Algorithm \ref{alg1}, we regard $R$ in (8) as the objective function $f({\bf{x}})$ in the Gibbs distribution formula (10), and denote the corresponding features as a $\left( {{K_u} + {K_d}+M} \right) \times 1$ binary vector $\bf{x}$. The first ${K_{\text{u}}} + {K_{\text{d}}}$ elements of ${\bf{x}}$ correspond to the uplink and downlink user scheduling results. The corresponding user is activated if ${x_i} = 1$ ($i \leqslant {K_u} + {K_d}$), and is kept inactivated if ${x_i} = 0$ ($i \leqslant {K_u} + {K_d}$). The following $M$ elements of ${\bf{x}}$ correspond to the antenna splitting results at the BS, i.e., the corresponding antenna will be configured to work in uplink mode if ${x_i} = 1$ ($i > {K_u} + {K_d}$), and downlink mode if ${x_i} = 0$ ($i > {K_u} + {K_d}$). The constraints in steps 3 and 4 of Algorithm \ref{alg1} correspond to the two constraints in (9) (i.e., $f({\bf{x}})$ should satisfy these constraints given in (9)). Besides, since the upper bounds of $\left| {{\cal K}_u^1} \right|$ and $\left| {{\cal K}_d^1} \right|$ are $\left| {{\cal M}_r^1} \right|$ and $\left| {{\cal M}_t^1} \right|$, we could not generate all the elements of ${\bf{x}}$ with Algorithm \ref{alg2} simultaneously. The generating procedure of ${\bf{x}}$ with Algorithm \ref{alg2} are divided into the following three steps: (1) Generate $N$ $K_u$- and $K_d$-dimensional multivariate Bernoulli random vectors ${{\bf{x}}_u}$ and ${{\bf{x}}_d}$ which satisfy $K_{\min }^u \leqslant {sum}({{\bf{x}}_u}) \leqslant {K_u}$, $K_{\min }^d \leqslant {sum}({{\bf{x}}_d}) \leqslant {K_d}$, respectively\footnote{$sum( \cdot )$ stands for the element summation of the multivariate Bernoulli random vectors, which equals to the number of users which are selected to be active.}. (2) Find ${\bf{x}}_u^1$ and ${\bf{x}}_d^1$ which have the smallest number of selected users among ${{\bf{x}}_u}$ and ${{\bf{x}}_d}$. Denote $K_u^1 = sum({\bf{x}}_u^1)$ and $K_d^1 = sum({\bf{x}}_d^1)$. (3) Generate $N$ $M$-dimensional multivariate Bernoulli random vectors ${{\bf{x}}_M}$ which satisfy $K_u^1 \leqslant {\text{sum}}({{\bf{x}}_M}) \leqslant M - K_d^1$. Then, the multivariate Bernoulli random vectors which need to be generated in step 3 of Algorithm \ref{alg1} are composed of ${\bf{x}}_u^1$, ${\bf{x}}_d^1$ and all the samples of ${\bf{x}}_M$. In this way, the selected uplink and downlink user sets keep unchanged among the $N$ new random samples. This is one approach to generate the multivariate Bernoulli random vectors with Algorithm 2 under the constraints in (9).

In addition to the joint antenna splitting and user scheduling problem in (\ref{equ9}), we also consider the following user scheduling problem only:
\begin{IEEEeqnarray}{rCl}\label{equ51}
	\mathop {\max }\limits_{{\cal K}_{\text{u}}^1 ,{\cal K}_{\text{d}}^ 1 }{\kern 10pt} && R  \IEEEyesnumber\IEEEyessubnumber*\\
	{\text{subject}}{\kern 5pt}{\text{to}}{\kern 10pt}&&K_{\min }^{\text{u}} \leqslant \left| {{\cal K}_{\text{u}}^1 } \right| \leqslant \left| {{{\cal M}_r}} \right|   \IEEEyessubnumber\\
	&&K_{\min }^{\text{d}} \leqslant \left| {{\cal K}_{\text{d}}^ 1 } \right| \leqslant \left| {{{\cal M}_t}} \right| \IEEEyessubnumber
\end{IEEEeqnarray}
where ${{\cal M}_r}$ and ${{\cal M}_t}$ are the specified uplink receiving antenna and downlink transmitting antenna sets, respectively. Similar to the optimization problem in (\ref{equ9}), ${\cal K}_{\text{u}}^{\text{1}}$ and ${\cal K}_{\text{d}}^{\text{1}}$ should also satisfy the corresponding constraints in (\ref{equadd0}). We further note that the feature $\bf{x}$ in Algorithm \ref{alg1} becomes a $\left( {{K_u} + {K_d}} \right) \times 1$ vector for the user scheduling problem in (20). Since the uplink and downlink antennas have been specified in this problem, there is no element in $\bf{x}$ that corresponds to the splitting results of antennas.

In practice, depending on the requirements of each user, the sets ${{\cal K}_u}$ and ${{\cal K}_d}$ may change over time. We note that Algorithm 2 can achieve the maximum SE for given sets of ${{\cal K}_u}$ and ${{\cal K}_d}$. Therefore, if ${{\cal K}_u}$ and ${{\cal K}_d}$ change, we just need to run Algorithm 2 with the new ${{\cal K}_u}$ and ${{\cal K}_d}$ to determine the new maximum SE that could be achieved in the network.

Finally, we note that in the numerical results discussed below, we continue the iterations in Algorithm \ref{alg1} until we obtain 100 sequential values of the objective function $R$, which satisfy the following stopping condition: all the absolute values of the difference between the former and the latter values are less than $10^{-6}$.

\section{Numerical Results} \label{sec:numerical}
In the numerical results, we assume that the radius of the picocell is 40 m, and the minimum distance between a user and the BS is 10 m. The carrier frequency is 2 GHz. The shadowing standard deviation is 3 dB for LOS links and 4 dB for NLOS links. We assume that all user-to-user channels experience NLOS links, and the shadowing standard deviation is 6 dB. The Rician factor of the SI channel $\kappa  = 1$, and the residual SI power $\sigma _{{\text{SI}}}^2 =  - 100$ dB. We assume $K_{\min }^{\text{u}} = K_{\min }^{\text{d}} = {K_{\min }}$ in the numerical results. The simulation results are averaged out over 1000 random channel realizations. All simulations were executed with Matlab R2018a on Windows 10 operating system equipped with Intel Core i7 CPU and 16 GB RAM.

In the performance evaluations, we call the process of solving (\ref{equ9}) with Algorithm \ref{alg1} as the Gibbs distribution based statistical joint antenna splitting and user scheduling (GS-J) algorithm, and describe the process of solving problem (\ref{equ51}) with Algorithm \ref{alg1} as the Gibbs distribution based statistical user scheduling (GS-U) algorithm. We compare the GS-U algorithm with the successive user selection (SUS) algorithm proposed in \cite{Ahn2016}, which considers only the user scheduling problem.

The optimal solution of the combinatorial optimization problem in (\ref{equ9}) can be obtained by exhaustive search.
In order to evaluate the performance of Algorithm \ref{alg1}, we also compare the solutions obtained with Algorithm \ref{alg1} and that achieved by exhaustive search. However, when the candidate antenna and user numbers are large, the sample space of the combinatorial optimization problem becomes very large and the complexity of exhaustive search becomes prohibitively high in practice. In other words, we could only execute the exhaustive search algorithm when either the antenna and user numbers are small or only the user scheduling problem in (\ref{equ51}) is addressed. Therefore, we consider two different settings in the numerical results:

\subsubsection{The Setting with Small Number of Antennas and Users}
In this setting, we assume that there are 6 antennas at the BS. There are 3 users requiring to transmit data to the BS, while another 3 users need to receive messages from the BS. In this setting, we perform exhaustive search for joint antenna splitting and user scheduling (ES-J) to solve (\ref{equ9}) and also exhaustive search for user scheduling only (ES-U) in (\ref{equ51}). In the numerical analysis of (\ref{equ51}), we set $M_r=2$ and $M_t=4$.

\subsubsection{The Setting with Large Number of Antennas and Users}
In this setting, we assume that there are 30 antennas at the BS. We further assume that there are 10 users requiring to transmit data to the BS, while another 10 users need to receive messages from the BS \footnote{The numbers of antennas and users we have used in the numerical analysis is not very large. However, the cardinality of the sample space, in this case, is ${2^{50}}$, which is large enough to show the advantage of our Gibbs distribution based statistical combinatorial optimization algorithm over exhaustive search. Moreover, our Gibbs distribution based statistical combinatorial optimization algorithm is also applicable when the antenna and user numbers are very large. On the other hand, for very large number of antennas at the BS, the advantage of joint antenna splitting and user scheduling becomes smaller compared to the scheme with only user scheduling (e.g. random antenna selection can obtain the approximately optimal subset of antennas if we specify the uplink and downlink antenna numbers first \cite{Lee2013}). Thus, joint antenna selection and user scheduling is not highly necessary in this case. However, our Gibbs distribution based statistical combinatorial optimization algorithm provides an efficient method for user scheduling to improve the SE of the communication system.}. In this setting, we perform exhaustive search for only user scheduling (ES-U) to solve (\ref{equ51}). Since problem (\ref{equ51}) has a much smaller sample space than (\ref{equ9}), it is easier to perform exhaustive search even when the antenna and user numbers are relatively large. In (\ref{equ51}), we set $M_r=10$ and $M_t=20$.

As described next, numerical results show that, when suitable parameters $\alpha$ and $\beta$ are used, the SE curves obtained with Algorithm \ref{alg1} overlap with the curves achieved with exhaustive search algorithms ES-J and ES-U for the optimization problems in (\ref{equ9}) and (\ref{equ51}), respectively.

\subsection{Simulation Results}
\begin{figure*}
	\centering
	\subfigure[$\eta=1$ and ${K_{\min }}=5$]{
		\label{fig3a} 
		\includegraphics[width=2.3in]{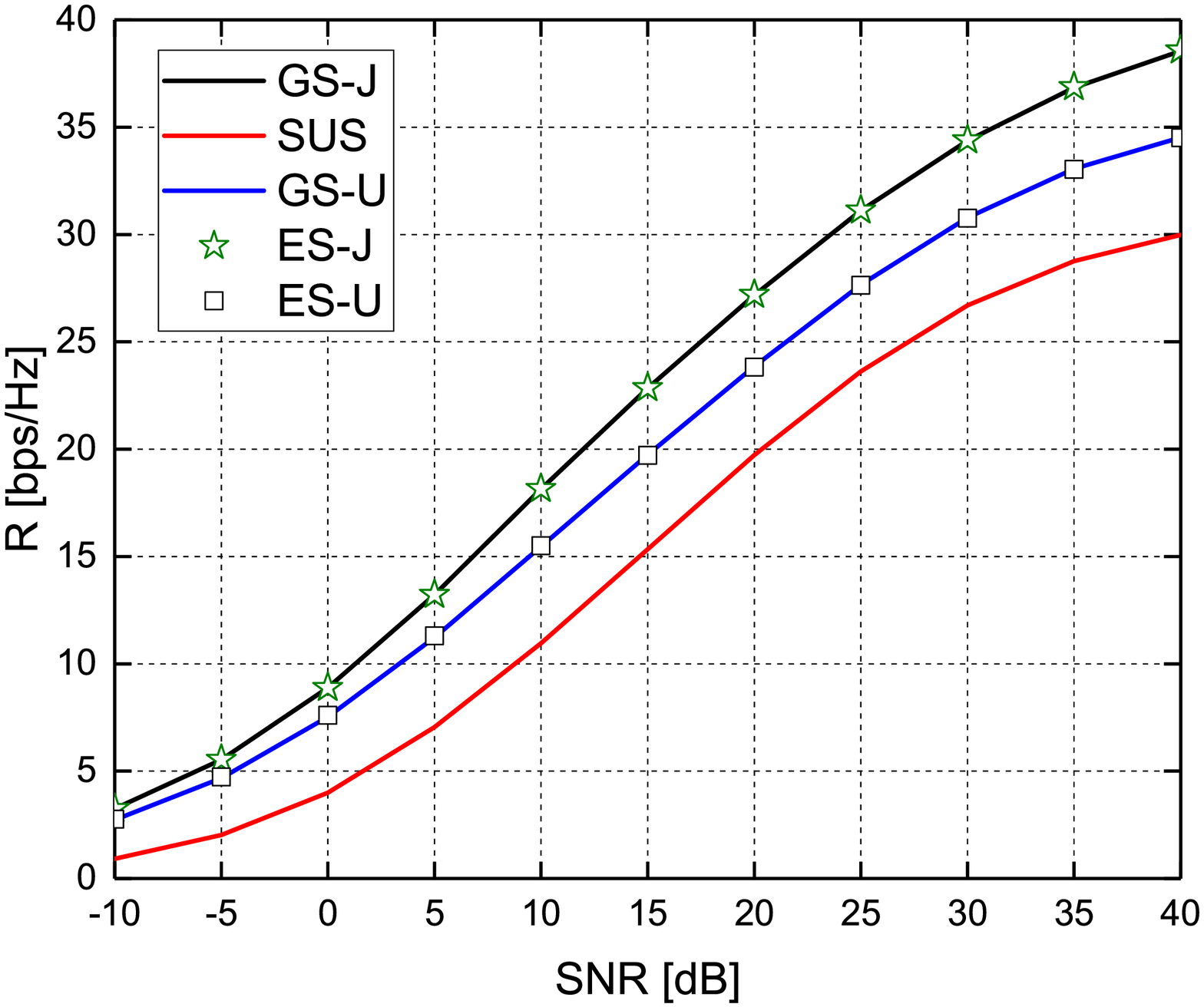}}
	\subfigure[SNR=20 dB and ${K_{\min }}=5$]{
		\label{fig3b} 
		\includegraphics[width=2.3in]{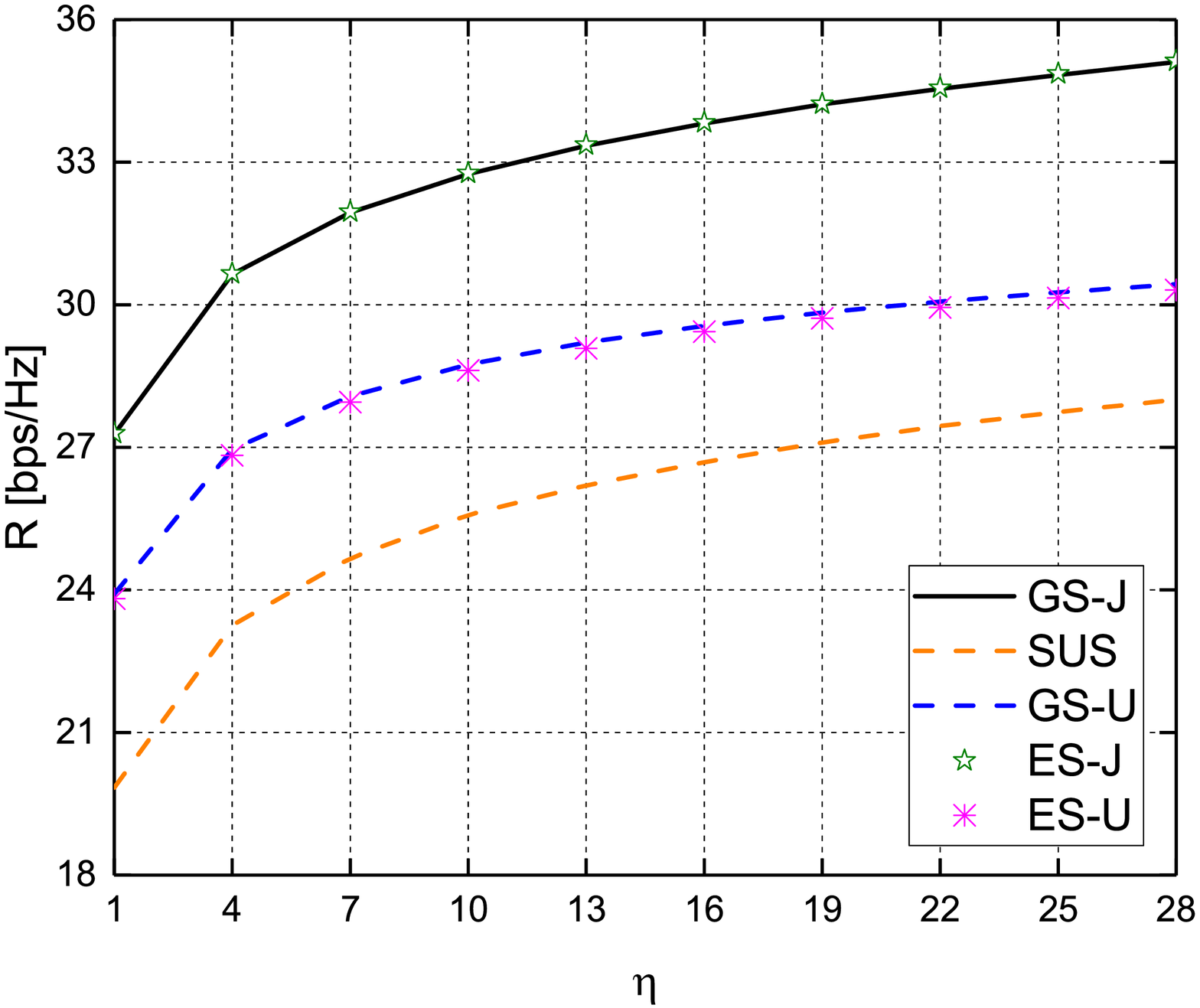}}
	\caption{Maximum SE achieved in a single-cell massive MIMO system with GS-J, GS-U, ES-J, ES-U and SUS algorithms, where $M=6$, ${N_u} = 3$, ${N_d} = 3$, $M_r=2$ and $M_t=4$.}
	\label{fig3} 
\end{figure*}

\begin{figure*}
	\centering
	\subfigure[$\eta=1$ and ${K_{\min }}=5$]{
		\label{fig2a} 
		\includegraphics[width=2.3in]{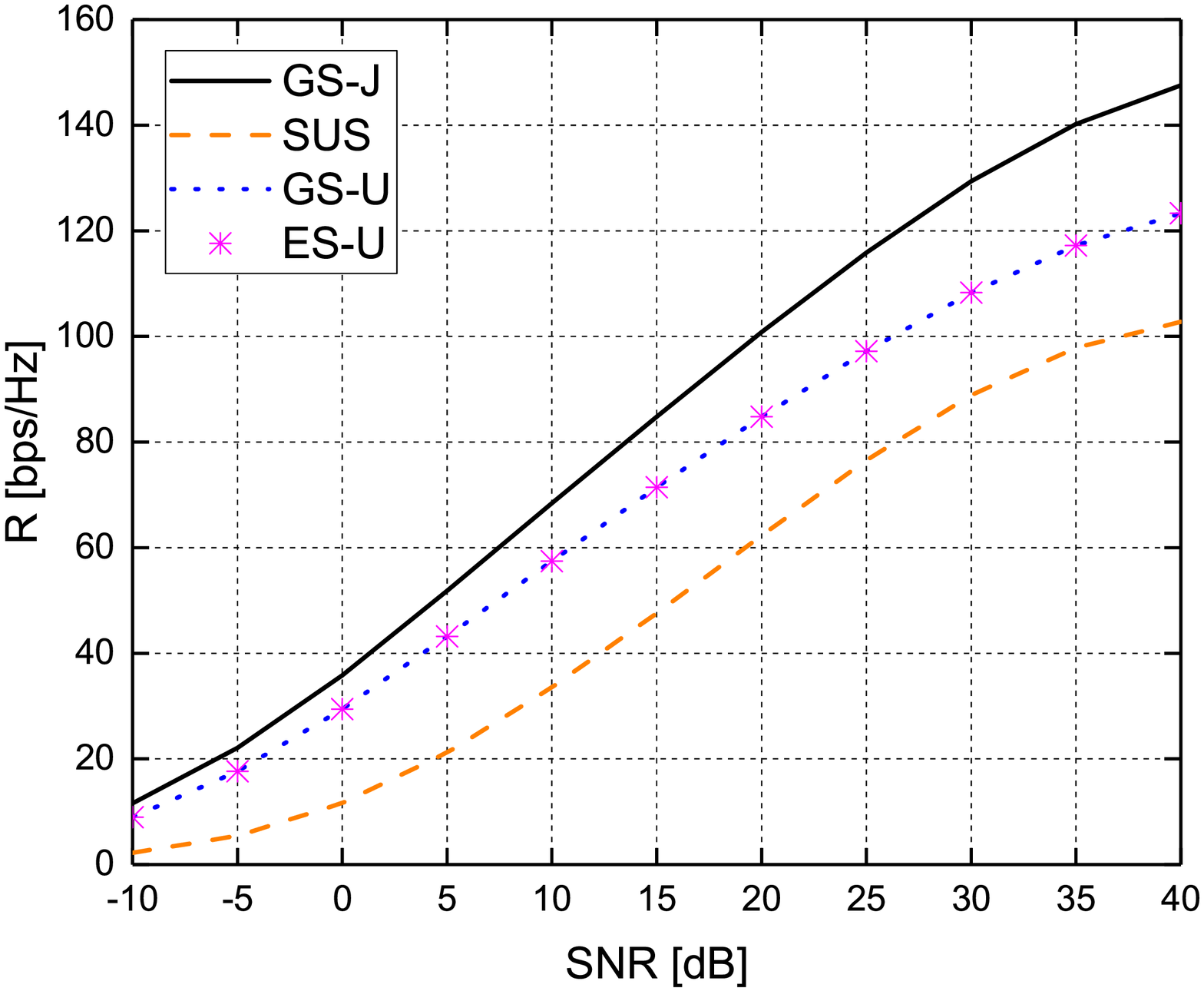}}
	\subfigure[SNR=20 dB and ${K_{\min }}=5$]{
		\label{fig2b} 
		\includegraphics[width=2.3in]{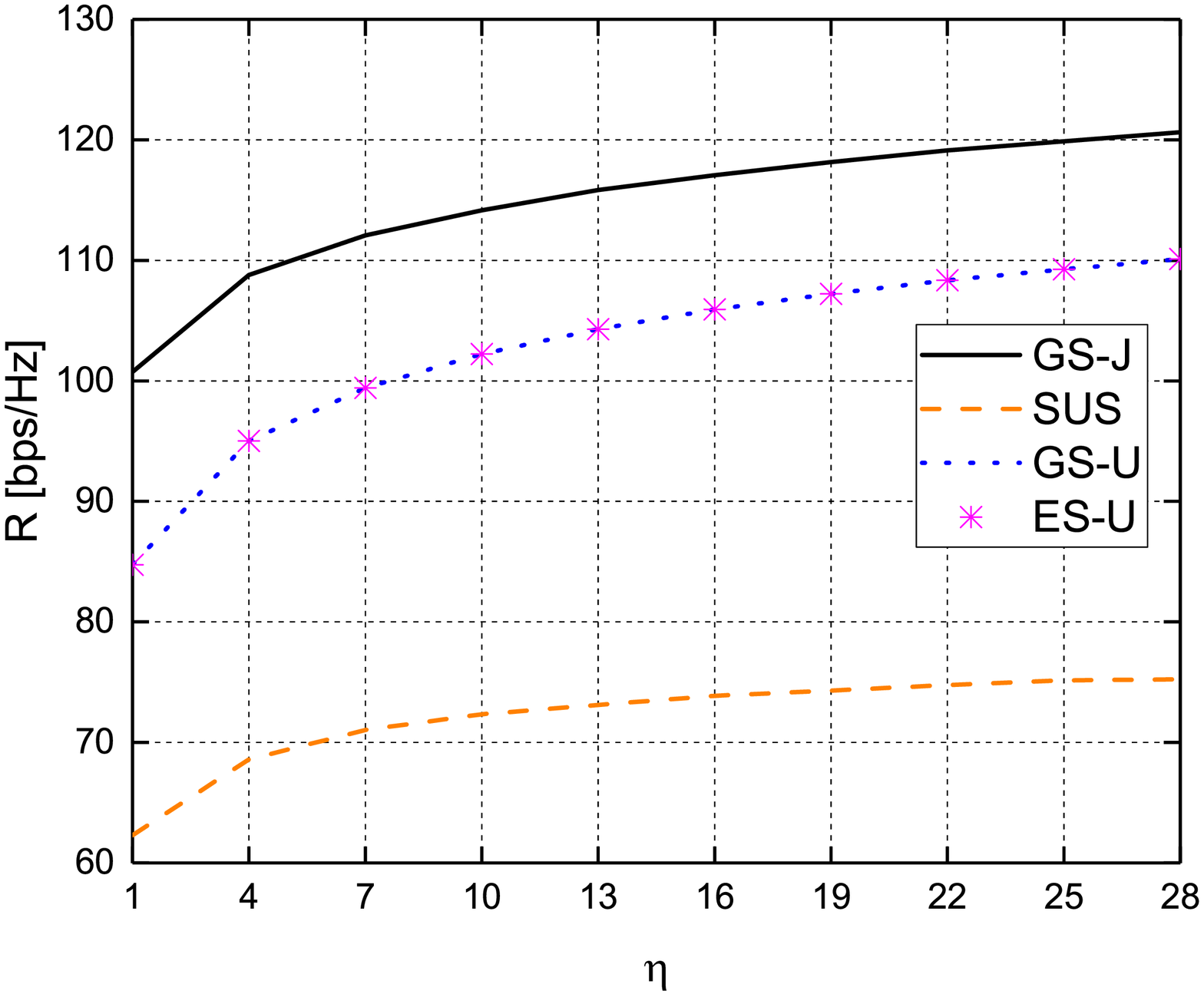}}
	\subfigure[SNR=20 dB and $\eta=1$]{
		\label{fig2c} 
		\includegraphics[width=2.3in]{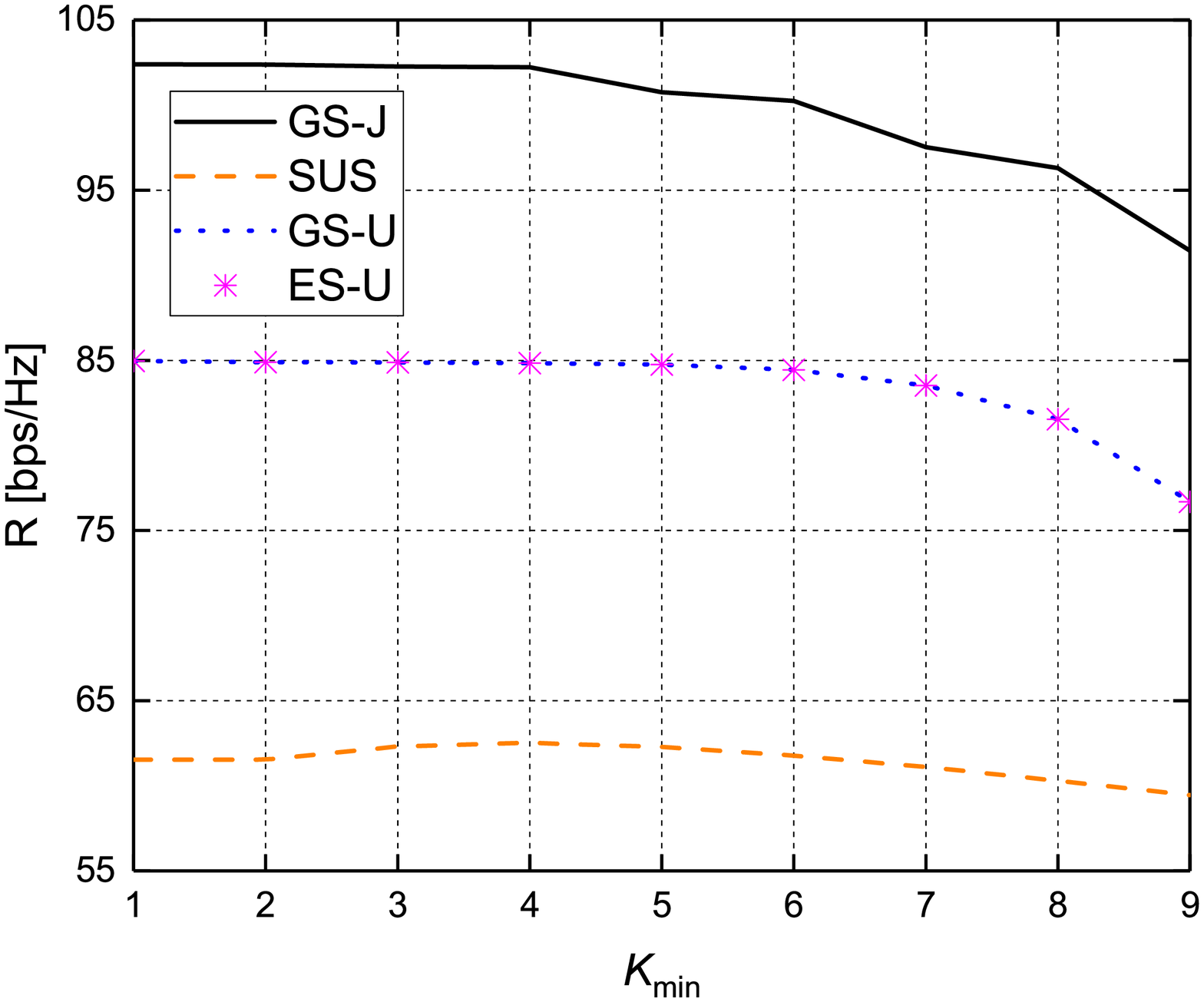}}
	\caption{Maximum SE achieved in a single-cell massive MIMO system with GS-J, GS-U, ES-U and SUS algorithms, where $M=30$, ${N_u} = 10$, ${N_d} = 10$, $M_r=10$ and $M_t=20$.}
	\label{fig2} 
\end{figure*}

Fig. \ref{fig3} and Fig. \ref{fig2} plot the simulation results of maximum SE achieved in single-cell massive MIMO systems as a function of key system parameters (e.g., uplink received SNR at the BS, power ratio $\eta  = {p_d}/{p_u}$, and minimum active user number ${K_{\min }}$). In Fig. \ref{fig3}, small number of antennas and users are considered whereas the setting with large number of antennas and users is addressed in Fig. \ref{fig2}. These figures show that GS-U algorithm can obtain the same performance as ES-U, and performs much better than the SUS algorithm. Additionally, GS-J that solves the joint antenna splitting and user scheduling problem can further improve the SE of the system. In Fig. 2, we also plot the curves of maximum SE achieved by ES-J, and these curves overlap with those achieved by GS-J.

In Fig. \ref{fig3a} and Fig. \ref{fig2a}, at relatively low SNR levels, since Gaussian noise is the dominant disturbance rather than interference in uplink signals, the maximum SE initially grows with increasing SNR. On the other hand, when SNR is larger than 25 dB, SI starts becoming the dominant disturbance component, and the rate of increase in SE begins to slow down. Indeed, as SNR tends to infinity, the Gaussian noise vanishes, and the maximum SE approaches an upper bound which is determined by the SI term.

In Fig. \ref{fig3b}, the GS-U and ES-U achieve more than 10 percent improvement compared to the SUS algorithm under the same SNR, $\eta$ and ${K_{\min }}$. Since ${p_{\text{d}}} = \eta{p_{\text{u}}}$, when we specify ${p_{\text{u}}}$, ${p_{\text{d}}}$ increases as $\eta$ increases, and therefore the maximum SE increases. However, when downlink transmit power grows, SI gets larger while the effect of the user-to-user interference diminishes, and the tendency of the maximum achieved SE in Fig. \ref{fig3b} is the result of these two factors. The tendency of the curves in Fig. \ref{fig2b} are similar with those in Fig. \ref{fig3b}. However, since the candidate antenna and user numbers become larger in Fig. \ref{fig2b}, the improvement between GS-U and SUS increases, compared to those in Fig. \ref{fig3b}.

In Fig. \ref{fig2c}, the maximum achieved SE with GS-U and ES-U algorithms is improved by more than 30 percent over that obtained by the SUS algorithm. When ${K_{\min }}$ is very small, the optimal subset of users may stay the same for different ${K_{\min }}$, so the curves in Fig. \ref{fig2c} keep constant at the beginning. As ${K_{\min }}$ increases, more users are selected, both SI and user-to-user interference grow, and the maximum achieved SE decreases. We also observe that, as before, highest performance results are attained with GS-J.

\begin{figure*}
	\centering
	\subfigure[]{
		\label{fig4a} 
		\includegraphics[width=2.3in]{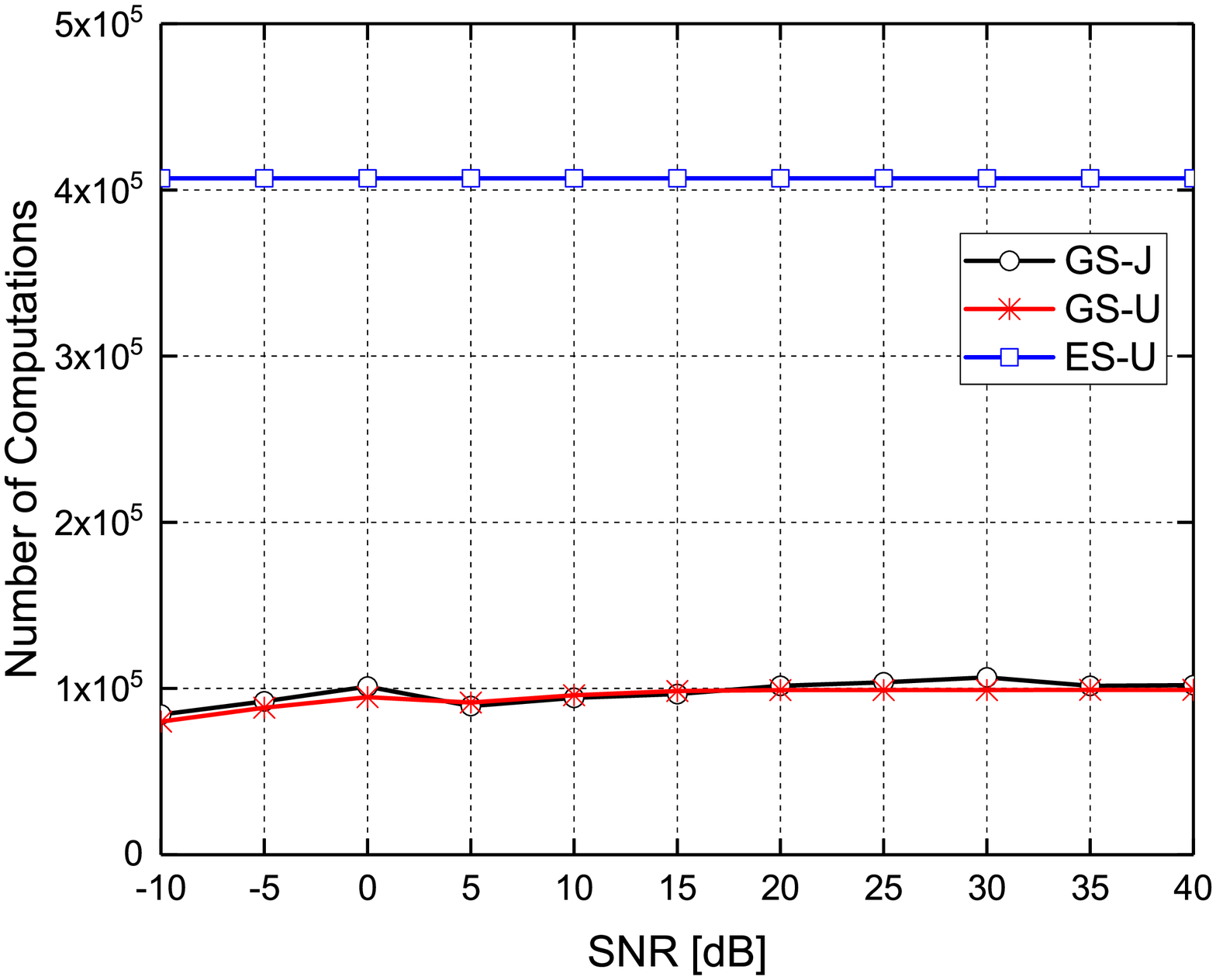}}
		\subfigure[]{
		\label{fig4b} 
		\includegraphics[width=2.3in]{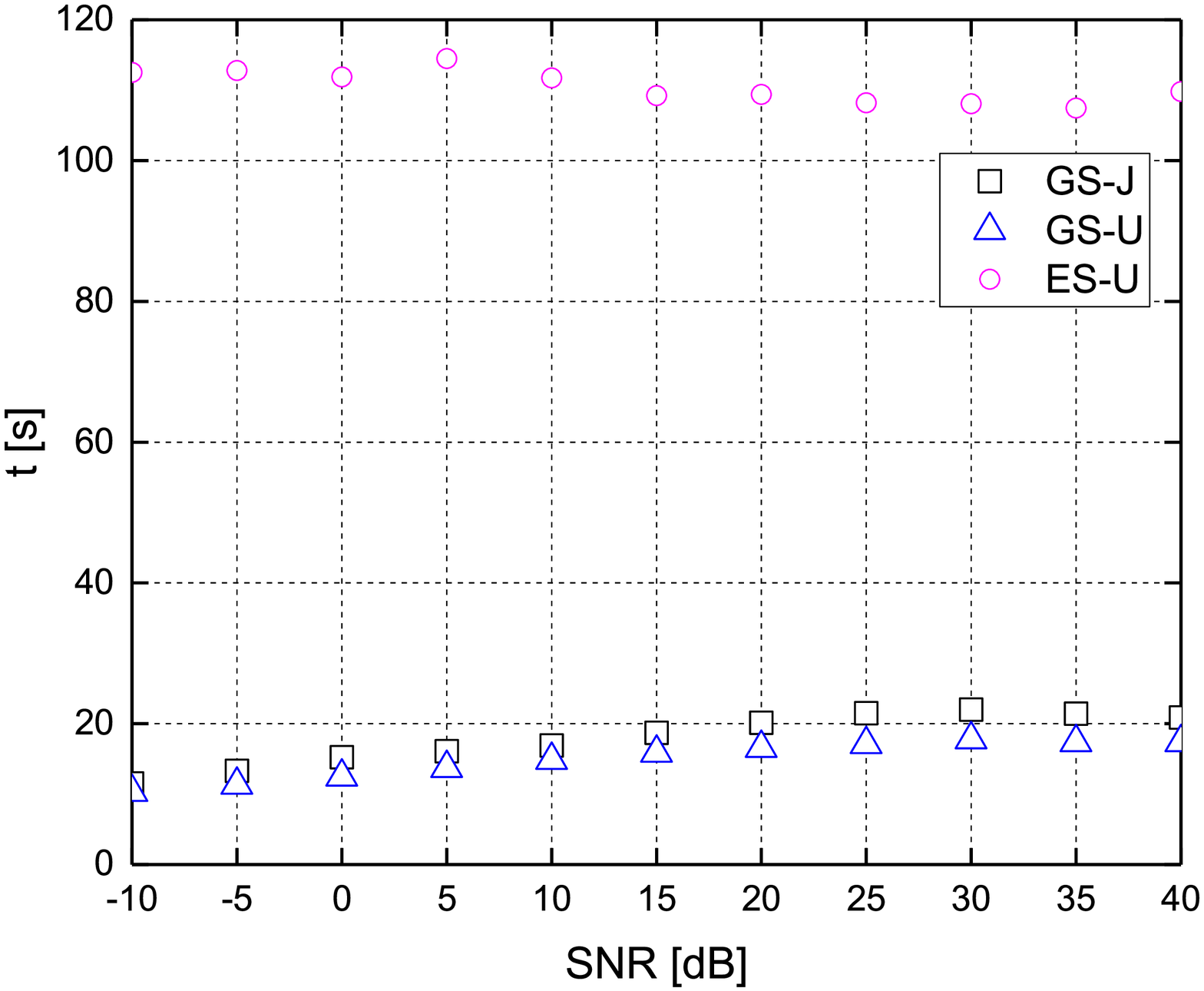}}
	\caption{Average number of computations of the objective function value and time consumed for the simulation of GS-J, GS-U, and ES-U algorithms, where $M=30$, ${N_u} = 10$, ${N_d} = 10$, $M_r=10$, $M_t=20$, $\eta=1$ and ${K_{\min }}=5$.}
	\label{fig4} 
\end{figure*}
Since Algorithm \ref{alg1} is a Gibbs distribution based statistical optimization method, it is difficult to analyze the computational complexity theoretically. Thus, we provide the computational complexity analysis based on numerical simulations. The computational complexity of Algorithm \ref{alg1} includes the generation of multivariate Bernoulli random vectors, checking of the constraints, and the calculation of $R$, while the complexity of the exhaustive search consists of only checking of the constraints and the calculation of $R$. However, numerical simulations show that in both Algorithm \ref{alg1} and exhaustive search, the time used for the calculation of $R$ dominates. Therefore, we use the number of times $R$ is calculated during the numerical simulations as a criterion for the computational complexity of Algorithm \ref{alg1} and exhaustive search. Besides, when all the numerical simulations are running with the same settings, average simulation time is another good measure for the computational complexity although it is hardware dependent.

Fig. 4 plots the curves for the computational complexity comparison of Algorithm \ref{alg1} and the exhaustive search method. The y-axis in Fig. \ref{fig4a} corresponds to the average number of times the objective function value $R$ is calculated during the numerical simulations. Fig. \ref{fig4a} shows that the complexity of GS-U is about 25\% of the ES-U algorithm, and the complexity of GS-J and GS-U are almost the same. However, the average complexity of ES-J is about $6.86861 \times {10^{13}}$, which is far greater than that of GS-J. Therefore, the complexity of solving the joint antenna splitting and user scheduling problem with Algorithm \ref{alg1} is much lower than that of exhaustive search method. Since $6.86861 \times {10^{13}}$ is too large and it would make it difficult to see the difference between GS-J and GS-U, we do not display it in Fig. \ref{fig4a}. Fig. \ref{fig4b} shows that the average time consumed by different algorithms. The runtimes of GS-J and GS-U are about 20\% of that of ES-U. The difference in the comparative percentages observed in Fig. \ref{fig4a} and Fig. \ref{fig4b} is due to ignoring the complexity of generating multivariate Bernoulli random vectors and checking the constraints in Fig. \ref{fig4a}. Since the runtime of the ES-J algorithm with large numbers of antennas and users is expected to be significantly long, ES-J was not tested during the simulations.

\section{Conclusion} \label{sec:conclusion}
A Gibbs distribution based statistical combinatorial optimization algorithm (Algorithm \ref{alg1}) for joint antenna splitting and user scheduling in full duplex massive MIMO systems is proposed in this paper. In order to verify the performance of Algorithm \ref{alg1}, we have also applied it to solve the user scheduling problem in single-cell massive MIMO systems, which has a much smaller sample space, compared with joint antenna splitting and user scheduling. Simulation results show that, with suitable parameters $\alpha$ and $\beta$, Algorithm \ref{alg1} can obtain the same performance as exhaustive search with much lower complexity. We have also demonstrated via this algorithm that the SE of the system can be further improved by joint antenna splitting and user scheduling.

\bibliographystyle{IEEEtran}
\bibliography{antenna_user}

\end{document}